\documentclass{iopart}

\usepackage[]{graphicx}

\begin{document}

\JPA

\title[Bridging the gap between correlation entropy functionals]{Bridging the gap between correlation entropy functionals in the mean spherical and the hypernetted chain approximations: a field theoretic description}

\author {Hiroshi Frusawa\footnote[1]{e-mail: frusawa.hiroshi@kochi-tech.ac.jp}}
\address{Laboratory of Statistical Physics, Kochi University of Technology, Tosa-Yamada, Kochi 782-8502, Japan.}

\begin{abstract}
The correlation entropy as a functional of radial distribution function $g(r)$ (or the total correlation function $h(r)=g(r)-1$) in classical fluids has been obtained from the second Legendre transform of the grand potential.
We focus on the correlation entropy difference between the two typical functionals in the mean spherical approximation (MSA) and the hypernetted chain (HNC) approximation.
While the entropy functional difference between these approximations is of a simple form, the diagrammatic approaches in the liquid state theory are quite different from each other.
Here we clarified the gap between the MSA and HNC functionals by developing a field theoretic description of the correlation functional theory that combines the variational principle of lower bound free energy, the conventional saddle-point approximation of a reference system to be optimized based on the variational principle, and the hybrid treatment of the saddle-point approximation and the fugacity expansion for modifying the primary optimization.
Our formulation demonstrates that the MSA functional is reproduced by the first maximization of the variational functional in the saddle-point approximation, and that the HNC functional is obtained from the improved maximization of the virial term due to the fugacity expansion around the MSA functional.
The virial term leads to the modification of a reference system interacting via the direct correlation function, thereby creating the correlation entropy difference.
\end{abstract}

\section{Introduction}

The liquid state theory (LST) has focused on investigating the constituent particle arrangement, or the particle-particle correlations in the atomic (or molecular) scale \cite{hansen}.
A wide variety of approximations for describing the structure in dense fluids has been proposed so far along the development of the LST [1-23].
In parallel, more than half of a century has passed since the free energy of classical fluids defined by the second Legendre transform was derived as a functional of density correlation field [24-28].
Focusing on the systematic treatment of the various approximations in the LST, the correlation functional theory (CFT) has been the most promising framework of the LST, because the CFT provides an integrated view on the diagrammatic method in the LST [1-28].
Furthermore, it is becoming an increasingly significant issue to develop the CFT that is able to consider fluctuations and heterogeneities of correlation fields as well as to encompass the LST, especially for glassy systems where out of equilibrium correlation fields are to be investigated over the meso-scale as well as the short-range scale [28-34].

The advantage of the CFT is that it is possible due to the diagrammatic formulation based on the second Legendre transform to systematically derive entropic contribution as a functional of density-density correlation function (see next section for the details), which will be referred to as correlation entropy [24-28].
The typical functionals of correlation entropy which we focus on are those given in the mean spherical approximation (MSA) and the hypernetted chain (HNC) approximation [1,24-28].
Recent theoretical studies have demonstrated that it matters which approximation of the LST is adopted, not only for improving the accuracy of thermodynamic quantities [1], but also for predicting phenomena in question, such as those in glass-forming liquids [33,34] and colloidal dispersions [35].

Based on quantitative discussions, it has been shown that it is essential either for evaluating the ideal glass transition points [33,34], or for explaining charge reversal of colloids due to counterions [35] to introduce a variety of empirical equations that covers a couple of approximations, including both of the HNC and MSA [33-36].
First of all, the correlation entropy in liquid state is directly related to the evaluation of both the Kauzmann's temperature in Lennard-Jones fluids [34] and the Kauzmann transition density for hard sphere fluids [33] at which the entropy difference between liquid and solid entropies (the configurational entropy) vanishes and the ideal glass transition occurs;
for instance, the most accurate results of the Kauzmann temperatures for Lennard-Jones fluids were obtained [34] using the Zerah and Hansen scheme [36] that continuously interpolates between the HNC and MSA.
Such a hybrid theory is also relevant to nonuniform fluids, such as electric double layer formed by counterions.
In the mixed approach of the so-called HNC/MSA [35], the colloid-counterion correlations are treated within the HNC approximation, whereas the MSA is used for counterion-counterion correlations.
It has been found useful to adopt the HNC/MSA when explaining overcharging phenomena that the number of condensed counterions exceeds the bare colloidal charge leading to charge reversal [35].
However, the underlying principle of these empirical treatments has not only been clarified, but also little attention has been paid to the bridging of the HNC and MSA using the perturbation theory, though there have been fundamental studies on the relationship between the HNC and MSA [4-7].

Our aim is thus to derive the correlation entropy difference between the entropy functionals given in the HNC and MSA using a field theoretic perturbation technique around the MSA.
In other words, we purpose to bridge the gap between the two approximations by revealing the systematic method to derive the entropy difference between the HNC and MSA functionals.
When we gain an aspect of systematically tuning the approximate forms between the HNC and MSA functionals, the field theoretic description could open up the possibility of the CFT that offers a seamless treatment of both the meso-scale fluctuations and heterogeneities of correlation fields, and the short-range correlations between particles going beyond the LST.

The remainder of this paper is organized as follows.
First, we describe the formal background of the above entropy difference formulated in the LST.
Next, we outline the two-step maximization of interaction potential based on the variational theory explained therein, in terms of general representations of the grand potential using configurational and functional integrals.
In section 4, the MSA functional of correlation entropy is derived by evaluating the grand potential in the saddle-point approximation and maximizing the variational functional with respect to the reference interaction potential.
In section 5, we develop a hybrid treatment of combining the saddle point approximation and the fugacity expansion, so that the perturbative contribution around the MSA functional due to the additional interaction potential can be evaluated field theoretically.
As a consequence, the improved maximization of the variational functional validates the correlation entropy difference between the MSA and HNC functionals of correlation entropy.
Lastly, the details of the two-step maximization are summarized, and the implications of our results are also discussed.

\section{Formal Background}

To investigate the particle-particle correlations in dense liquids, it is necessary to find the behaviors of the total correlation function $h(r)$ (or the radial distribution function $g(r)\equiv 1+h(r)$) and the direct correlation function $c(r)$, in addition to the bare potential $\beta v(r)\, (\beta\equiv1/k_BT)$ of the original system at distance $r$ in the unit of thermal energy $k_BT$.
The LST has yielded various approximate forms of $c(r)$, using the set of two equations between $c(r)$ and $h(r)$:
$c(r)$ of uniform liquids having the smeared density of $\overline{\rho}$ is defined by the Ornstein-Zernike (OZ) equation [1],
 \begin{equation}
h(r)=c(r)+\overline{\rho}\int d{\bf r}' 
c(|{\bf r}-{\bf r}'|)h(r'),
   \label{oz}
   \end{equation}
while satisfying the closure relation,
\begin{equation}
   g(r)=e^{-\beta v(r)-c(r)+h(r)-b(r)},
   \label{closure}
\end{equation}
where the bridge function $b(r)$ denotes the sum of elementary diagrams \cite{hansen}.

Let $\Xi$ and $\beta\Omega\{\beta v\}=-\ln \Xi$ be respectively the grand partition function and the associated thermodynamic potential which is referred to as the grand potential.
The above closure relation is equivalent to the following identity,
\begin{eqnarray}
\frac{\delta(\beta \Gamma/V)}{\delta g}=-\frac{\overline{\rho}^2}{2}\beta v(r),
\label{identity}
\end{eqnarray}
regarding the correlation functional $\Gamma\{g\}$, or the second Legendre transform of $\Omega$:
\begin{eqnarray}
\frac{\beta\Gamma\{g\}}{V}&=&\frac{\beta\Omega\{\beta v\}}{V}-\frac{\overline{\rho}^2}{2}\int dr\,g(r)\beta v(r)\nonumber\\
\frac{\delta(\beta\Omega/V)}{\delta(\beta v)}&=&\frac{\overline{\rho}^2}{2}g(r),
\label{secondL}
\end{eqnarray}
with $V$ denoting the system volume.
Equations (\ref{identity}) and (\ref{secondL}) have formed the basis of the CFT [24-28].

The correlation functional $\Gamma\{g\}$ is related to the correlation entropy functional $s$ per unit volume.
The functionals in the MSA and HNC approximation are respectively given as follows [1-33]:
\begin{eqnarray}
\left(-\frac{s}{k_B}\right)_{\mathrm{MSA}}&=&\frac{\beta\Gamma_{\mathrm{MSA}}\{g\}}{V}+\overline{\rho}\ln z
\nonumber\\
&=&\overline{\rho}\ln\overline{\rho}-\overline{\rho}-\frac{1}{2}\int dk
\left[\ln\{1+\overline{\rho} h(k)\}-\overline{\rho} h(k)
\right]
\label{msa_entropy}
\end{eqnarray}
and
\begin{eqnarray}
\left(-\frac{s}{k_B}\right)_{\mathrm{HNC}}&=&\frac{\beta\Gamma_{\mathrm{HNC}}\{g\}}{V}+\overline{\rho}\ln z\nonumber\\
&=&\overline{\rho}\ln\overline{\rho}-\overline{\rho}
+\frac{\overline{\rho}^2}{2}\int dr\left\{g(r)\ln g(r)-h(r)\right\}\nonumber\\
&&\quad-\frac{1}{2}\int dk
\left[\ln\{1+\overline{\rho} h(k)\}-\overline{\rho} h(k)+\frac{\overline{\rho}^2h^2(k)}{2}
\right]
\label{hnc_entropy}
\end{eqnarray}
where $z$ represents the fugacity.
This paper focuses on the correlation entropy difference, $\Delta s/k_B=(s/k_B)_{\mathrm{HNC}}-(s/k_B)_{\mathrm{MSA}}$, which reads
\begin{eqnarray}
-\frac{\Delta s}{k_B}&=&\frac{\beta\Gamma_{\mathrm{HNC}}\{g\}}{V}-\frac{\beta\Gamma_{\mathrm{MSA}}\{g\}}{V}\nonumber\\
&=&\frac{\overline{\rho}^2}{2}\int dr\left\{g(r)\ln g(r)-h(r)-\frac{h^2(r)}{2}\right\}.
\label{difference}
\end{eqnarray}
Notably, eq. (\ref{difference}) is of a simple form though the MSA and the HNC approximation are quite different from each other in terms of the diagrammatic method as mentioned below.

In the HNC approximation, we use the approximate closure relation,
\begin{equation}
   g(r)=e^{-\beta v(r)-c(r)+h(r)},
   \label{hnc}
\end{equation}
ignoring the bridge function $b(r)$ [1].
In terms of the CFT, the identity (\ref{identity}) leads to eq. (\ref{hnc}) by discarding the two-particle irreducible diagrams of $\Gamma$ [24-28]. 
The primary MSA, on the other hand, is given by $g(r)=0$ for $r<\sigma_{\mathrm{eff}}$ and $\beta v(r)+c(r)=0$ for $r\geq\sigma_{\mathrm{eff}}$ with $\sigma_{\mathrm{eff}}$ denoting the effective diameter [1-3], and one of its generalizations provides the soft MSA [4-9]:
  \begin{equation}
   \int dr\,g(r)\left\{
   \beta v(r)+c(r)\right\}=0.
   \label{closure}
   \end{equation}
The solution of the OZ equation (\ref{oz}) using such kind of the MSA has not only been obtained as the analytic forms of $c(r)$ in various systems including hard sphere fluids in the Percus-Yevick (PY) approximation and Coulomb fluids, but has also been demonstrated to provide better results than the HNC approximation in some systems such as hard sphere fluids [1-3].
In this paper, we will use the MSA as the therm that covers this type of approximations including the basic MSA, the soft MSA, and the PY approximation [1-23].
It is straightforward to validate the MSA based on the optimized random phase approximation (RPA) of the thermodynamic perturbation theory; however, a preset reference system is needed in the optimized RPA [19-23].
In addition, the MSA, or the optimized RPA, includes the resummation of diagrams, which is not along the diagrammatic treatment of the HNC [1-28].
We thus have had a gap between the two approximations of the MSA and HNC, even though the hybrid treatment of these approximations has been found useful [13-17].

\section{Two-step maximization: outline in terms of general formulations}
\subsection {Our strategy}
To derive $\Delta s$ given by eq. (\ref{difference}), essential use will be made of the following three methods: (i) lower bound variational principle [1,8], other than the well-known upper bound approach referred to as the thermodynamic perturbation theory [1], (ii) the saddle-point approximation, and (iii) a field-theoretic description of the fugacity expansion [37-40].
We introduce a grand potential as a variational functional of arbitrary two-body interaction potential to be optimized, relying not on the usual upper bound due to the Gibbs-Bogoliubov inequality, but on the lower bound [1,8] of the true grand potential $\Omega\{\beta v\}$.
Setting that $\beta v=u+w$, we have a grand potential $\Omega\{u\}$ of a reference system interacting via unknown interaction potential $u$. The addition of a residual interaction energy provides the lower bound of the true grand potential $\Omega\{\beta v\}$ such that [1,8]
 \begin{eqnarray}
  \frac{\beta\mathcal{L}\{u;g\}}{V}&\equiv&\frac{\beta\Omega\{u\}}{V}+\frac{\overline{\rho}^2}{2}\int dr g(r)w(r)\leq
 \frac{\beta\Omega\{\beta v\}}{V}.
   \label{start} 
   \end{eqnarray}
Maximization of the variational functional $\mathcal{L}\{u;g\}$ with respect to $u$ yields
   \begin{eqnarray}
   \left.
   \frac{\beta\mathcal{L}\{u;g\}}{\delta u}
   \right|_{u=u^*}&=&0\nonumber\\
   \frac{\beta\mathcal{L}\{\beta u^*;g\}}{V}&=&
   \frac{\beta\Gamma\{g\}}{V}+\frac{\overline{\rho}^2}{2}\int dr g(r)\beta v(r),
\label{def_L}
   \end{eqnarray}
where the correlation functional $\Gamma\{g\}$ corresponds to the second Legendre transform of $\Omega\{u\}$:
   \begin{eqnarray}
\frac{\beta\Gamma\{g\}}{V}&=&\frac{\beta\Omega\{u^*\}}{V}-\frac{\overline{\rho}^2}{2}\int dr g(r)u^*(r)\nonumber\\
\left.
\frac{\delta(\beta\Omega/V)}{\delta u}\right|_{u=u^*}&=&\frac{\overline{\rho}^2}{2}g(r)
\label{def_gamma}
\end{eqnarray}
using an external source $u$ instead of the true potential $\beta v$ in eq. (\ref{secondL}).

There are two steps to reach the maximized functional of $\mathcal{L}\{u;g\}$ given by eq. (\ref{def_L}).
The first maximization of $\mathcal{L}\{u_1;g\}$ ($\beta v=u_1+w_1$) will be performed using the RPA functional of $\Omega\{u_1\}$ obtained from the saddle point approximation, similarly to the optimized RPA [19-23].
It is to be noted that we need not to presume a reference system such as hard sphere fluids, other than the optimized RPA, and that the maximization determines an appropriate system of particles interacting via the optimized potential $u_1^*$ that mimics the original state.
Accordingly, we will have $\Gamma_{\mathrm{MSA}}$, or the correlation entropy in the MSA given by eq. (\ref{msa_entropy}).
We will further consider the variational functional $\mathcal{L}\{u_1^*+u_2;g\}$ to improve the first maximization using a field theoretic description of the fugacity expansion around the MSA functional, so that the HNC functional will be obtained.

\subsection{Configurational integral representation}

In transforming configurational integral representation to functional integral form, it is convenient to introduce an instantaneous density correlation function $G_{\hat{\rho}}$ that
\begin{eqnarray}
   G_{\hat{\rho}_N}({\bf x},{\bf y})=\hat{\rho}_N({\bf x})\hat{\rho}_N({\bf y})
   -\hat{\rho}_N({\bf x})\delta(|{\bf x}-{\bf y}|),
\label{bare_correlation}
   \end{eqnarray}
using the instantaneous density defined by
\begin{equation}
\hat{\rho}_N({\bf x})=\sum_{i=1}^N\delta({\bf x}-{\bf x}_i).
\label{bare_density}
\end{equation}
The density correlation function $G_{\hat{\rho}_N}({\bf x},{\bf y})$ allows us to express the bare interaction energy $U\{G\{\hat{\rho}_N\};\beta v\}$ as follows:
\begin{eqnarray}
\beta U\{G_{\hat{\rho}_N};\beta v\}&=&\frac{1}{2}\int d{\bf x}\int d{\bf y}\,
G_{\hat{\rho}_N}({\bf x},{\bf y})\beta v(|{\bf x}-{\bf y}|).
\label{u-start}
\end{eqnarray}
The grand potential $\Omega\{\beta v\}$ of a true particle system interacting via the bare interaction potential $\beta v$ then reads
\begin{eqnarray}
  e^{-\beta\Omega\{\beta v\}}&=&\mathrm{Tr}\,\exp\left[-\beta U\{G_{\hat{\rho}_N};\beta v\}\right]\nonumber\\
  \mathrm{Tr}&\equiv&
  \sum_{N=0}^{\infty}\frac{z^{N}}{N!}\,\int d{\bf x}_1\cdots\int d{\bf x}_N,
\label{f-start}
\end{eqnarray}
using the configurational integral of $N$-particle positions, $\{{\bf x}_1,\cdots,{\bf x}_N\}$.
For uniform liquids, the ensemble average of the instantaneous density correlation function, $\left<G_{\hat{\rho}_N}(r)\right>$, is related to the radial distribution function $g(r)$ such that
\begin{eqnarray}
\frac{\delta(\beta\Omega/V)}{\delta(\beta v)}
&=&\frac{1}{2V}\left[
\frac{\mathrm{Tr}\,G_{\hat{\rho}_N}\,e^{-\beta U\{G_{\hat{\rho}_N};\beta v\}}}{ e^{-\beta\Omega\{\beta v\}}}
\right]
\nonumber\\
&=&\frac{1}{2V}\left<G_{\hat{\rho}_N}(r)\right>=\frac{\overline{\rho}^2}{2}g(r),\nonumber\\
\end{eqnarray}
which proves eq. (\ref{secondL}).

\subsection{The first step: maximization of variational functional in the saddle-point approximation}

Dividing the interaction potential as $v=u_1+w_1$, the variational functional for the first maximization is given by
\begin{eqnarray}
\frac{\beta\mathcal{L}\{u_1;g\}}{V}&=&\frac{\beta\Omega\{u_1\}}{V}+\frac{\overline{\rho}^2}{2}\int dr g(r)w_1(r).
\end{eqnarray}
In the first maximization, we take the saddle-point approximation of functional integral form for evaluating the grand potential $\beta\Omega\{u_1\}$ of a reference particle system interacting via an unknown interaction potential $u_1(r)$.
The functional integral form can be incorporated into eq. (\ref{f-start}) using the following identity that maps the instantaneous density $\hat{\rho}_N({\bf x})$ to the density field $\rho({\bf x})$ : 
\begin{eqnarray}
  1&=&\int\,D\rho\,\prod_{\{{\bf x}\}}\>\delta\left[
   \rho({\bf x})-\hat{\rho}_N({\bf x})
   \right]\nonumber\\
&=&\int\int\,D\rho\,D\psi\,e^{\int d{\bf x}i\psi({\bf x})
   \left\{\hat{\rho}_N({\bf x})-\rho({\bf x})\right\}},
\label{rho-identity}
\end{eqnarray}
where a one-body potential field $\psi$, dual to $\rho$, is also introduced in the Fourier transformed form.

Combining eqs. (\ref{f-start}) and (\ref{rho-identity}), we have
\begin{eqnarray}
e^{-\beta\Omega\{u_1\}}&=&\int\int\,D\rho\,D\psi\,
e^{-\beta U\{G_{\rho};u_1\}-\int d{\bf x}i\rho({\bf x})\psi({\bf x})}\,\mathrm{Tr}\,e^{\int d{\bf x}i\hat{\rho}_N({\bf x})\psi({\bf x})}\nonumber\\
&=&\int D\psi\,
e^{-\beta\mathcal{H}_1\{u_1;\psi\}},
\label{h1}
\end{eqnarray}
where the delta functional allows $U\{G_{\hat{\rho}_N};u_1\}$ to be replaced by $U\{G_{\rho};u_1\}$.
We implement the Gaussian integration over the density field in eq. (\ref{h1}), extracting the density fluctuation, $\Delta\rho({\bf x})=\rho({\bf x})-\overline{\rho}$, around the smeared density $\overline{\rho}$ with the separation of $U\{G_{\rho};u_1\}=U\{G_{\overline{\rho}};u_1\}+U\{G_{\Delta\rho};u_1\}$.
It follows that 
\begin{eqnarray}
\fl\quad\beta\mathcal{H}_1\{u_1;\psi\}=U\{G_{\overline{\rho}};u_1\}+
\frac{1}{2}\int d{\bf x}\int d{\bf y}\,\psi({\bf x})u_1^{-1}(|{\bf x}-{\bf y}|)\psi({\bf y})
+\int d{\bf x}\,i\overline{\rho}\psi-ze^{i\psi({\bf x})}.\nonumber\\
\label{h1_phi}
\end{eqnarray}
As detailed below, the saddle point approximation of the functional integral over the potential field $\psi({\bf x})$ yields the RPA functional.
It will be further verified that the first maximization of $\beta\Omega\{u_1\}$ with respect to $u_1$ provides the MSA functional given by eq, (\ref{msa_entropy}):
\begin{eqnarray}
\frac{\beta\mathcal{L}\{-c;g\}}{V}&=&
\frac{\beta\Gamma_{\mathrm{MSA}}\{g\}}{V}+\frac{\overline{\rho}^2}{2}\int dr g(r) \beta v(r)\nonumber\\
u^*_1(r)&=&-c(r),
   \end{eqnarray}
implying that the soft core system of liquid particles interacting via minus the direct correlation function $-c(r)$ is assigned to the reference system that has been optimized within the first maximization.

\subsection{The second step: improved maximization using the fugacity expansion around the MSA functional}

Next, we consider an improved reference system interacting via the sum of the additional potential $u_2$ and the above reference potential $u_1=-c$ determined by the first maximization:
\begin{equation}
v=-c+u_2+w_2.
\label{second_division}
\end{equation}
The additional potential $u_2$ is to be optimized for further maximizing the variational functional difference due to the perturbation, which reads
\begin{eqnarray}
\frac{\beta\mathcal{L}\{-c+u_2;g\}}{V}&=&\frac{\beta\Omega\{-c+u_2\}}{V}+\frac{\overline{\rho}^2}{2}\int dr g(r)w_2(r)\nonumber\\
&=&\frac{\beta\mathcal{L}\{-c;g\}}{V}+\frac{\beta\Delta\Omega\{u_2\}}{V}-\frac{\overline{\rho}^2}{2}\int dr g(r)u_2(r)\nonumber\\
\Delta\Omega\{u_2\}&=&\Omega\{-c+u_2\}-\Omega_{\mathrm{MSA}}\nonumber\\
\frac{\beta\Omega_{\mathrm{MSA}}}{V}&=&\frac{\beta\Gamma_{\mathrm{MSA}}\{g\}}{V}-\frac{\overline{\rho}^2}{2}\int dr g(r) c(r).
\label{second_variational}
\end{eqnarray}
Maximizing the variational functional $\mathcal{L}\{-c+u_2;g\}$ with respect to $u_2$, we have
\begin{eqnarray}
\mathcal{L}\{-c+u^*_2;g\}&=&\mathcal{L}\{-c;g\}+\Delta\Gamma\{g\}\nonumber\\
\frac{\beta\Delta\Gamma\{g\}}{V}&=&\frac{\beta\Delta\Omega\{u_2^*\}}{V}-\frac{\overline{\rho}^2}{2}\int dr g(r)u^*_2(r)
\label{delta_gamma}
\end{eqnarray}
using the optimized potential determined by the maximum condition: $\delta(\beta\Delta\Omega)/\delta u_2|_{u_2=u_2^*}=\overline{\rho}^2g/2$.

In the improved maximization, we take a hybrid form due to the different treatment of $-c$ and $u_2$ in transforming into the functional integral representation as follows:
\begin{eqnarray}
e^{-\beta\Omega\{-c+u_2\}}&=&\int\int\,D\rho\,D\psi\,
e^{-\beta U\{G_{\rho};-c\}-\int d{\bf x}i\rho({\bf x})\psi({\bf x})}\nonumber\\
&&\qquad\qquad\times \mathrm{Tr}\,e^{-\beta U\{G_{\hat{\rho}_N};u_2\}+\int d{\bf x}i\hat{\rho}_N({\bf x})\psi({\bf x})}\nonumber\\
&=&\int D\psi\,
e^{-\beta\mathcal{H}_2\{-c+u_2;\psi\}}.
\label{h2}
\end{eqnarray}
We can see a difference between eqs. (\ref{h1}) and (\ref{h2}) that $U\{G_{\hat{\rho}_N};-c+u_2\}=U\{G_{\rho};-c\}+U\{G_{\hat{\rho}_N};u_2\}$ leaving the term $U\{G_{\hat{\rho}_N};u_2\}$.
In eq. (\ref{h2}), we integrate the density fluctuation field $\Delta\rho({\bf x})$ as before, giving
\begin{eqnarray}
\fl\>\beta\mathcal{H}_2\{-c+u_2;\psi\}=\frac{-1}{2}\int d{\bf x}\int d{\bf y}\,\psi({\bf x})c^{-1}(|{\bf x}-{\bf y}|)\psi({\bf y})
+\int d{\bf x}i\overline{\rho}\psi({\bf x})-\ln\Xi,
\label{h2_original}
\end{eqnarray}
where
\begin{eqnarray}
\Xi=\mathrm{Tr}\,e^{-\beta U\{G_{\hat{\rho}_N};u_2\}+\int d{\bf x}i\hat{\rho}_N({\bf x})\psi({\bf x})}.
\label{xi_def}
\end{eqnarray}
It will be found that $e^{-\beta\Omega_{\mathrm{MSA}}}=\int D\psi\,e^{-\beta\mathcal{H}_1\{-c;\psi\}}$, so that the additional functional $\Delta\Omega\{u_2\}$ given by eq. (\ref{second_variational}) may read
\begin{eqnarray}
e^{-\beta\Delta\Omega\{u_2\}}
&=&\frac{\int D\psi\,
e^{-\beta\mathcal{H}_2\{-c+u_2;\psi\}}}{\int D\psi\,
e^{-\beta\mathcal{H}_1\{-c;\psi\}}}.
\label{deltaomega}
\end{eqnarray}
We then perform the improved maximization of $\Delta\Omega\{u_2\}$ using the fugacity expansion around the MSA functional based on the hybrid form of eqs. (\ref{h2}) and (\ref{deltaomega}).
As a consequence, it will be verified that  $\beta\Delta\Gamma\{g\}/V=-\Delta s/k_B$ (see below eq. (\ref{delta_gamma2})):
the virial term arising from the above fugacity expansion explains the difference between the MSA and HNC entropies.

\section{The MSA functional in the first maximization}
\subsection{Performing the saddle-point approximation}

In the saddle-point approximation of eq. (\ref{h1}), the saddle-point (or the mean-field) equation, $\delta\mathcal{H}_1/\delta\psi|_{\psi=i\psi^*}=0$, reads
\begin{eqnarray}
\ln\left\{\frac{\rho^*({\bf x})}{z}\right\}&=&\int d{\bf y} u_1(|{\bf x}-{\bf y}|)\{
\rho^*({\bf y})-\overline{\rho}\}\nonumber\\
\rho^*({\bf x})&=&z\exp \{-\psi^*({\bf x})\},
\label{mf}
\end{eqnarray}
as found from the expression (\ref{h1_phi}) of $\mathcal{H}_1$.
Expanding $\mathcal{H}_1$ around the saddle-point field $\psi^*$,  we have
\begin{eqnarray}
&&\beta\mathcal{H}_1\{u_1;i\psi^*+\phi\}
=\frac{1}{2}\int d{\bf x}\int d{\bf y}\,\phi({\bf x})m(|{\bf x}-{\bf y}|)\phi({\bf y})\nonumber\\
&&\quad+\beta U\{G_{\overline{\rho}};u_1\}
+\int d{\bf x}
\left\{\rho^*({\bf x})\ln\rho^*({\bf x})-\rho^*({\bf x})-\rho^*({\bf x})\ln z\right\}\nonumber\\
&&m(|{\bf x}-{\bf y}|)\equiv u_1^{-1}(|{\bf x}-{\bf y}|)+\overline{\rho}\delta({\bf x}-{\bf y})
\label{h1_sp}
\end{eqnarray}
for $\psi=i\psi^*+\phi$ in the saddle-point approximation that $ze^{i\psi({\bf x})}=z\,e^{-\psi^*}(1+i\phi-\phi^2/2)$ regarding the last term on the right hand side of eq. (\ref{h1_phi});
actually, the solution of the saddle-point equation (\ref{mf}) is $\rho^*({\bf x})=\overline{\rho}=z$ (or $\psi^*=0$) in the absence of an external field.
The Gaussian integration over the $\phi$-field thus yields
\begin{eqnarray}
\fl\quad\frac{\beta\Omega\{u_1\}}{V}&=&
\frac{1}{2}\int dr\,G_{\overline{\rho}}(r)u_1(r)+\frac{1}{2}\sum_{k}\ln\{1+\overline{\rho}u_1(k)\}+\overline{\rho}\ln\overline{\rho}-\overline{\rho}-\overline{\rho}\ln z,
\label{omega_u1}
\end{eqnarray}
where $G_{\overline{\rho}}(r)=\overline{\rho}^2-\overline{\rho}\delta(r)$.

\subsection{Maximization with respect to the interaction potential of reference system}

The second derivative of $\mathcal{L}\{u_1\}$ with respect to $u_1$ is of the following form:
\begin{eqnarray}
\frac{\delta^2(\beta\mathcal{L}\{u_1\}/V)}{\delta u_1^2}=
-\frac{\overline{\rho}^2}{2\{1+\overline{\rho}u_1(k)\}^2}\leq0,
\end{eqnarray}
where the left hand side is due to the second term on the right hand side of eq. (\ref{omega_u1}).
The above negativity of the second derivative assures that the maximum of the variational functional $\mathcal{L}\{u_1\}$ is found from the stability condition at the extremum $u_1^*$ [4-8]:
\begin{eqnarray}
\left.
\frac{\delta(\Omega\{u_1\}/V)}{\delta u_1}\right|_{u_1=u_1^*}
=\frac{1}{2}\left\{
\frac{-\overline{\rho}^2u_1^*(k)}{1+\overline{\rho}u_1^*(k)}
\right\}
+\frac{\overline{\rho}^2}{2}=\frac{\overline{\rho}^2}{2}g(k),
\end{eqnarray}
which reads
\begin{eqnarray}
h(k)
=\frac{-u_1^*(k)}{1+\overline{\rho}u_1^*(k)}=-\frac{1}{m^*(k)}
\label{first-solution}
\end{eqnarray}
using the definition of $m^*$ given by eq. (\ref{h1_sp}).
Because eq. (\ref{first-solution}) is identical to the Fourier transformed expression of the OZ equation (\ref{oz}) setting that $u_1^*=-c$, eq. (\ref{first-solution}) further implies that
\begin{eqnarray}
m^*(|{\bf x}-{\bf y}|)&=&-c^{-1}(|{\bf x}-{\bf y}|)+\overline{\rho}\delta(|{\bf x}-{\bf y}|)\nonumber\\
&=&-h^{-1}(|{\bf x}-{\bf y}|).
\label{msa_propagator}
\end{eqnarray}

Let us now go back to eq. (\ref{omega_u1}) with $u_1$ set to be $u_1^*=-c$, reminding that $c(0)=\int dk\,c(k)$,
\begin{eqnarray}
\frac{\overline{\rho}^2}{2}\int dr\,h(r)c(r)+\frac{\overline{\rho}}{2}\int dk\,c(k)
=\frac{\overline{\rho}}{2}\int dk\,h(k),
\end{eqnarray}
and the relation $1+\overline{\rho}h(k)=\{1-\overline{\rho}c(k)\}^{-1}$ due to the OZ equation (\ref{oz}).
It follows that
\begin{eqnarray}
\frac{\beta\Gamma\{-c\}}{V}&=&\frac{\beta\Omega\{-c\}}{V}+\frac{\overline{\rho}^2}{2}\int drg(r)c(r)\nonumber\\
&=&\overline{\rho}\ln\overline{\rho}-\overline{\rho}-\overline{\rho}\ln z
-\frac{1}{2}\sum_{k}\left[\ln\{1+\overline{\rho}h(k)\}-\overline{\rho}h(k)\right].
\label{gamma_1}
\end{eqnarray}
Comparison of eqs. (\ref{msa_entropy}) and (\ref{gamma_1}) demonstrates that the correlation entropy of the MSA functional is exactly reproduced by $\Gamma\{-c\}$.

\section{The HNC functional due to the improved maximization}
\subsection{Performing the fugacity expansion around the MSA functional}

In evaluating $\Xi$ given by eq. (\ref{xi_def}), we truncate the fugacity expansion of the "Tr" operator up to the second order [37-40]:
\begin{eqnarray}
\fl\qquad\quad\Xi=1+z\int d{\bf x}_1\,e^{i\psi({\bf x}_1)}
+\frac{z^2}{2}\int d{\bf x}_1\int d{\bf x}_2\,e^{-u_2(|{\bf x}_1-{\bf x}_2|)+i\psi({\bf x}_1)+i\psi({\bf x}_2)}.
\end{eqnarray}
As a consequence, the expression (\ref{h2_original}) of $\mathcal{H}_2$ is put into a tractable form:
\begin{eqnarray}
\mathcal{H}_2\{-c+u_2;\psi\}=\mathcal{H}_1\{-c;\psi\}
+\frac{z^2}{2}\int d{\bf x}_1\int d{\bf x}_2\,f\{\psi\}
\label{h2_app}
\end{eqnarray}
using the approximation,
\begin{eqnarray}
-\ln\Xi&\approx&-z\int d{\bf x}_1\,e^{i\psi({\bf x}_1)}
+\frac{z^2}{2}\int d{\bf x}_1\int d{\bf x}_2f\{\psi\}\nonumber\\
f\{\psi\}&=&e^{\int d{\bf x}\,i\psi(x)\hat{\rho}_2({\bf x})}\left\{
1-e^{-u_2(|{\bf x}_1-{\bf x}_2|)}\right\},
\label{h2_xi}
\end{eqnarray}
where two particle density, $\hat{\rho}_2({\bf x})=\sum_{i=1}^2\delta({\bf x}-{\bf x}_i)$, has been introduced.
Equations (\ref{h2_app}) and (\ref{h2_xi}) allow us to perform the hybrid treatment of the saddle-point approximation and the fugacity expansion:
we treat the right hand side of eq. (\ref{h2_app}) as a perturbation to $H_1\{-c;\psi\}$  while applying the saddle-point approximation to $H_1\{-c;\psi\}$ as well as eq. (\ref{h1_sp}).
Accordingly, we have the additional grand potential $\Delta\Omega$ of the following form:
\begin{eqnarray}
\beta\Delta\Omega\{u_2\}&=&
\frac{\overline{\rho}^2}{2}\int d{\bf x}_1\int d{\bf x}_2\,\left<f\{\phi\}\right>\nonumber\\
\left<f\{\phi\}\right>&=&
\frac{\int D\phi\,f\{\phi\}e^{-\frac{1}{2}\int d{\bf x}\int d{\bf y}\phi({\bf x})m^*(|{\bf x}-{\bf y}|)\phi({\bf y})}}
{\int D\phi\,e^{-\frac{1}{2}\int d{\bf x}\int d{\bf y}\phi({\bf x})m^*(|{\bf x}-{\bf y}|)\phi({\bf y})}}.
\label{deltaomega2}
\end{eqnarray}
As seen from Appendix for the details, eq. (\ref{deltaomega2}) reads
\begin{eqnarray}
\left<f\{\phi\}\right>
=e^{h(|{\bf x}_1-{\bf x}_2|)}\left\{1-e^{-u_2(|{\bf x}_1-{\bf x}_2|)}\right\},
\label{virial}
\end{eqnarray}
where use has been made of eq. (\ref{msa_propagator}) and the relation $z^2f\{i\psi^*+\phi\}=\overline{\rho}^2f\{\phi\}$.
These expressions (\ref{deltaomega2}) and (\ref{virial}) are similar to the second virial term [1,37-40], other than the prefactor $e^h$ in eq. (\ref{virial}).

\subsection{Maximization with respect to the additional interaction potential}

The variational functional, $\mathcal{L}\{-c+u_2;g\}$, given by eq. (\ref{second_variational}) adds eq. (\ref{deltaomega2}) to the first variational functional of $\mathcal{L}\{-c;g\}$.
Correspondingly, the second derivative of  $\mathcal{L}\{-c+u_2;g\}$ with respect to $u_2$ is due to the $u_2$-dependence of $\left<f\{\phi\}\right>$ given by eq. (\ref{virial}):
\begin{eqnarray}
\frac{\delta^2(\beta\Delta\Omega\{u_2\}/V)}{\delta u_2^2}=
-\frac{\overline{\rho}^2}{2}e^{h(r)-u_2(r)}\leq0,
\label{inequality2}
\end{eqnarray}
assuring the negativity of the second derivative.
Equation (\ref{inequality2}) implies that the variational functional, $\mathcal{L}\{-c+u_2^*;g\}$, is maximized at the extremum $u_2^*$ given by
\begin{eqnarray}
\left.
\frac{\delta(\beta\Delta\Omega)}{\delta u_2}\right|_{u_2=u_2^*}=\frac{\overline{\rho}^2}{2}e^{h(r)-u^*_2(r)}=\frac{\overline{\rho}^2}{2}g(r),
\label{improve1}
\end{eqnarray}
which reads
\begin{equation}
u^*_2(r)=h(r)-\ln g(r).
\label{improve2}
\end{equation}
Plugging eq. (\ref{improve2}) into eq. (\ref{delta_gamma}), we have
\begin{eqnarray}
\frac{\beta\Delta\Gamma}{V}
&=&\frac{\overline{\rho}^2}{2}\int dr\,\left\{e^{h(r)}-g(r)\right\}+\frac{\overline{\rho}^2}{2}\int dr\,\{g(r)\ln g(r)-g(r)h(r)\}\nonumber\\
&\approx&\frac{\overline{\rho}^2}{2}\int dr\,\left\{g(r)\ln g(r)-h(r)-\frac{h^2(r)}{2}\right\},
\label{delta_gamma2}
\end{eqnarray}
using the expansion, $e^h\approx1+h+h^2/2$, in the last line of eq. (\ref{delta_gamma2}).
Comparing eqs. (\ref{difference}) and (\ref{delta_gamma2}), it is verified that
\begin{equation}
\frac{\beta\Delta\Gamma}{V}=-\frac{\Delta s}{k_B}=\frac{\beta\Gamma_{\mathrm{HNC}}}{V}-\frac{\beta\Gamma_{\mathrm{MSA}}}{V}.
\label{delta_gamma2}
\end{equation}
We have thus demonstrated that the virial term arising from the fugacity expansion up to the second order creates the correlation entropy difference between the MSA and HNC functionals.

\section{Summary and discussion}

\begin{table}
\caption{Comparison between the conventional liquid state theory and our field theoretical description based on the variational principle (variational field theory). We provide four aspects of the correlation functionals of our concern, the constraint on the radial distribution function, two approximations (approx.) in deriving the correlation functionals, and the difference between the two approximations. The related equations are also indicated for comparison.}
\begin{center}
\includegraphics[width=10.5cm]{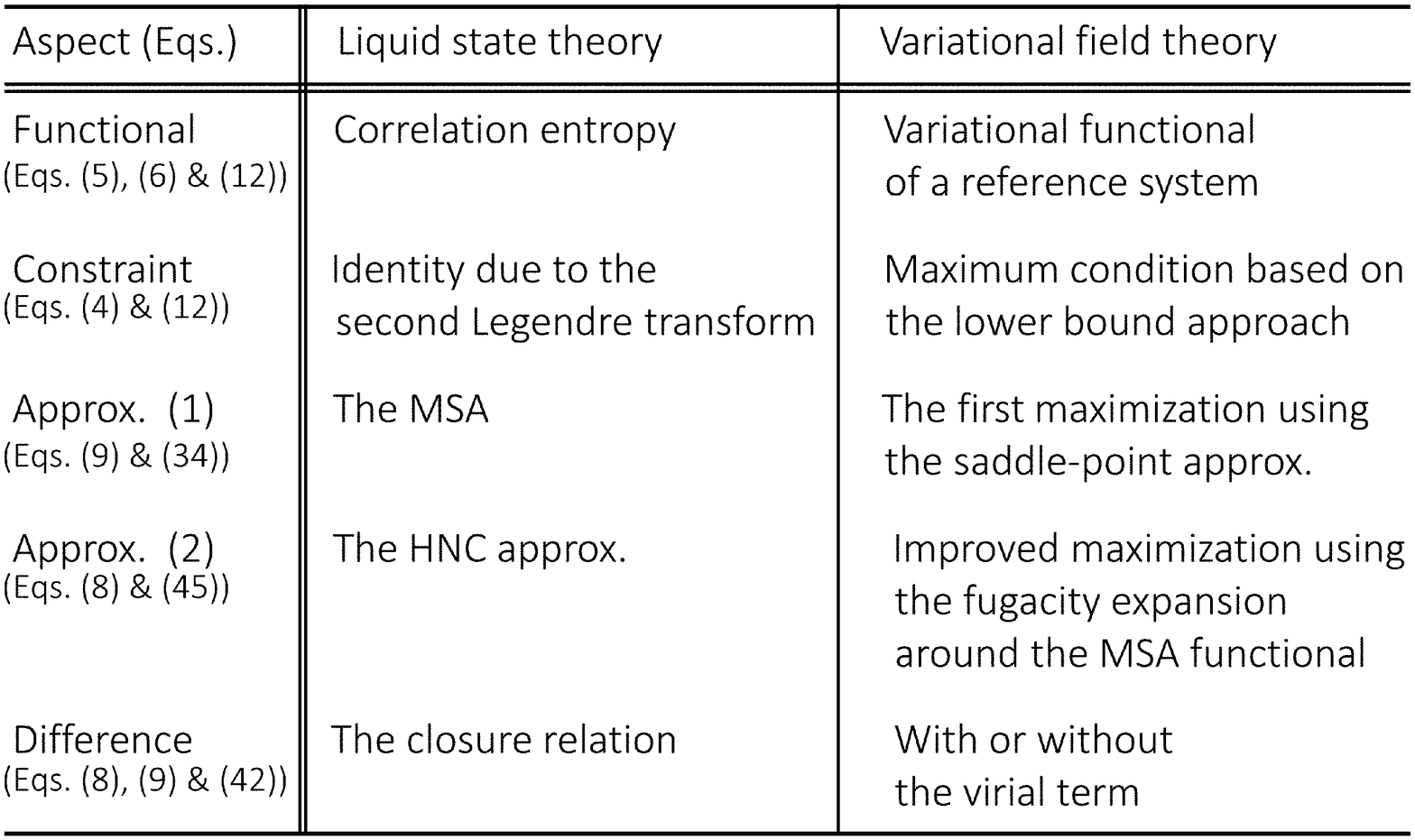}
\end{center}
\end{table}

In Table 1, the relationship between the LST and our field theoretical description based on the variational principle is summarized especially in terms of the correlation entropy difference between the MSA and HNC functionals.
Table 1 lists four aspects of correspondences: (i) the correlation functionals to be focused on, (ii) the basis of the constraint on the radial distribution function for being related to the interaction potential, (iii) two kinds of approximations in evaluating the functionals, and  (iv) the reason for the correlation entropy difference due to the different approximations.
Table 1 also provides the equations to be compared as well as the details of the four correspondences.

In general, the connection of our field theoretical description to the LST is clarified by focusing on the correlation functional of $\Gamma\{g\}$ that is directly related to the correlation entropy such as eqs. (\ref{msa_entropy}) and (\ref{hnc_entropy}).
While the second Legendre transform of the grand potential yields the correlation functional $\Gamma\{g\}$ in the LST, the maximization of variational functional as the lower bound yields the equivalent of the second Legendre transform in our formulation.
As seen from Table 1, the correlation entropy difference between the MSA and HNC functionals is simply explained by the perturbative term around the MSA functional, according to our field theoretical description, though the MSA is quite different from the HNC in terms of the LST.
The details follow:  the MSA functional of the correlation entropy is reproduced by the correlation functional $\Gamma\{g\}$ that has been obtained from combining the saddle-point approximation of a reference grand potential and its maximization with respect to an interaction potential characterizing the reference system, whereas the HNC functional is given not only by adding to the MSA functional the virial term due to the field theoretical perturbation, but also by improving the maximization of the grand potential with respect to an additional interaction potential.

Going back to eq. (\ref{delta_gamma2}), it is found that the virial term itself is not necessarily equated with the correlation entropy difference (\ref{difference}).
In transforming $\Omega$ to $\Gamma$, eq. (\ref{delta_gamma}) shows that the interaction energy $(\overline{\rho}^2/2)\int dr\,g(r)u^*_2(r)$ is subtracted using the optimized potential of $u_2^*=\ln g-h$, which is one of the essential terms for explaining the correlation entropy difference.
In addition, the expansion $e^h\approx 1+h+h^2/2$ in the virial term is also needed.
In other words, the summary given in Table 1 needs to be supplemented for bridging the gap between the correlation entropy functionals in the MSA and the HNC approximation; the meaning of $u_2^*$ is to be clarified.
Since the HNC discards the bridge term in the closure relation (\ref{hnc}) as mentioned in the introduction, it follows from eqs. (\ref{hnc}) and (\ref{improve2}) that
\begin{eqnarray}
u^*_2&=&h-\ln g=v+c_{\mathrm{HNC}}=\Delta c\nonumber\\
\Delta c
&=&c_{\mathrm{HNC}}-c_{\mathrm{MSA}},
\label{direct_difference}
\end{eqnarray}
for $r\geq \sigma_{\mathrm{eff}}$ in the case of the original MSA, because the direct correlation function $c_{\mathrm{MSA}}$ in the original MSA satifies $v+c_{\mathrm{MSA}}=0\,(r\geq \sigma_{\mathrm{eff}})$ while the HNC retain the term: $v+c_{\mathrm{HNC}}\neq 0$ [1].
Equation (\ref{direct_difference}) implies that the improved maximization due to consideration of the virial term corrects the direct correlation function outside the core, thereby creating the correlation entropy difference.

We can also regard the reference system as an optimized system that appropriately mimics the original system. From the aspect of the mimicked system, there is a large difference between the short-range behaviors of particle-particle correlations in the two systems interacting via $u_1^*=-c$ in the first maximization and $u_1^*+u_2^*=-c+h-\ln g$ including the improved potential.
The first maximization reproducing the MSA functional validates that the mimicked particles have soft cores, or the interaction potential having a finite value even inside the core.
On the other hand, the improved maximization recovering the HNC functional adds the harshly repulsive core over the range satisfying $g(r)=0\,(r<\sigma_{\mathrm{eff}})$.
It is further to be noted that our formulation verifies the HNC apprxoimation using the expansion $e^h=1+h+h^2/2$, which is invalid for the inside of the core as well as the contact state at $r=\sigma_{\mathrm{eff}}$.
It remains to be investigated that the new approximation without the expansion of $e^h$ provide any meaningful results in the shor-range scale, going beyond the LST.

A significant aspect of our achievement is revealed by eq. (\ref{direct_difference}).
In this paper, we have demonstrated that the virial term added to the MSA explains the difference between the approximate closures of the MSA and HNC in terms of the perturbation theory.
Our formulation makes it possible to further improve the approximate forms of the direct correlation function systematically.
Therefore, extension of the present variational approach to the correlation field theory using functional integral representation [41] could open up the possibility of simultaneously and seamlessly treating the meso-scale fluctuations and inhomogeneities of correlation fields out of equilibrium, and the short-range correlations between particles going beyond the LST.
The variational theory developed herein is a starting point to formulate such a field theoretic framework, which is increasingly becoming significant especially for glass-forming liquids where out-of-equilibrium correlations as well as the correlation entropy are to be investigated [28-34].
Furthermore, it is straightforward to extend the present formulation to that for nonuniform liquids, which is a promising route to improve the HNC/MSA for overcharging phenomena (mentioned at the beginning) in an alternative way without the use of the LST, or integral equation theories [35].

\section*{Acknowledgments}
We are grateful to R. Akiyama and K. Koga for helpful discussions.

\section*{Appendix. Derivation of the virial term given by eqs. (\ref{deltaomega2}) and (\ref{virial})}

Combining eqs. (\ref{deltaomega}) and (\ref{h2_app}), we have
\begin{eqnarray}
e^{-\beta\Delta\Omega\{u_2\}}
&=&\frac{\int D\psi\,
e^{-\beta\mathcal{H}_2\{-c+u_2\psi\}}}
{\int D\psi\,
e^{-\beta\mathcal{H}_1\{-c;\psi\}}}.\nonumber\\
&=&
\frac{\int D\psi\,
e^{-\mathcal{H}_1\{-c;\psi\}
-\frac{z^2}{2}\int d{\bf x}_1\int d{\bf x}_2\,f\{\psi\}}}
{\int D\psi\,
e^{-\beta\mathcal{H}_1\{-c;\psi\}}}.
\label{appendix_deltaomega}
\end{eqnarray}
Expanding $H_1$ around the solution of the saddle-point equation (\ref{mf}) in the numerator of eq. (\ref{appendix_deltaomega}) as well as in the denominator, we have
\begin{eqnarray}
\fl\qquad
e^{-\beta\Delta\Omega\{u_2\}}
=\frac{\int D\phi\,e^{-\frac{1}{2}\int d{\bf x}\int d{\bf y}\,\phi({\bf x})m^*(|{\bf x}-{\bf y}|)\phi({\bf y})
-\frac{\overline{\rho}^2}{2}\int d{\bf x}_1\int d{\bf x}_2\,f\{\phi\}}}
{\int D\phi\,e^{-\frac{1}{2}\int d{\bf x}\int d{\bf y}\,\phi({\bf x})m^*(|{\bf x}-{\bf y}|)\phi({\bf y})}},
\end{eqnarray}
which reads
\begin{eqnarray}
e^{-\beta\Delta\Omega\{u_2\}}
&=&\left<
e^{-\frac{\overline{\rho}^2}{2}\int d{\bf x}_1\int d{\bf x}_2\,f\{\phi\}}
\right>\nonumber\\
&\approx&
e^{
-\frac{\overline{\rho}^2}{2}\int d{\bf x}_1\int d{\bf x}_2\,\left<f\{\phi\}\right>
},
\label{average1}
\end{eqnarray}
according to eq. (\ref{deltaomega2}).

In eq. (\ref{deltaomega2}), the average of $\left<f\{\phi\}\right>$ implies that
\begin{eqnarray}
\left<f\{\phi\}\right>
&=&\left<e^{\int d{\bf x}\,i\phi({\bf x})\hat{\rho}_2({\bf x})}\right>\left\{1-e^{-u_2(|{\bf x}_1-{\bf x}_2|)}\right\}.
\label{average2}
\end{eqnarray}
The average on the right hand side of eq. (\ref{average2}) is reduced to the Gaussian integration over the $\phi$-field with eq. (\ref{msa_propagator}) rewritten as $(m^*)^{-1}=-h$.
It follows that
\begin{eqnarray}
\left<f\{\phi\}\right>
&=&e^{\frac{1}{2}\int d{\bf x}\int d{\bf y}\hat{\rho}_2({\bf x})h({\bf x}-{\bf y})\hat{\rho}_2({\bf y})}
\left\{1-e^{-u_2(|{\bf x}_1-{\bf x}_2|)}\right\},
\label{average3}
\end{eqnarray}
thereby proving eq. (\ref{virial}).
Equations (\ref{average1}) and (\ref{average3}) thus verify eqs. (\ref{deltaomega2}) and (\ref{virial}).


\end{document}